\documentclass[aps,twocolumn,pra,showpacs]{revtex4}
\usepackage[dvips]{epsfig}

\date{\today}

\begin{document}

\title{A narrow plane cut near the crystal surface increases the probability of capture into the stable channeling
motion up to 99 percent}

\author{V.~V.~Tikhomirov \footnote {vvtikh@mail.ru}}
\address
{Research Institute for Nuclear Problems, Belarus State
University, Bobruiskaya 11, 220030 Minsk, Belarus}

\draft
\begin{abstract}
It is shown that a narrow plane cut near the crystal surface
considerably increases the probability of capture into the stable
channeling motion of positively charged particles entering a
crystal at angles smaller than a quarter of the critical
channeling angle with respect to the crystal planes. At smallest
incidence angles the capture probability reaches 99 percent. A
pair of crystals with cuts bent in orthogonal planes allows to
reach a 99.9 percent efficiency of single-pass deflection of a
proton beam with an ultra small divergence. Conditions necessary
for efficient single-pass deflection of protons from the LHC beam
halo are discussed.

\end{abstract}
 \maketitle

The phenomenon of positively charged particle channeling in
crystals has already found several applications in high energy
physics. In particular, the deflection of channeling protons and
ions by bent crystals is widely used for beam extraction from
high-energy accelerators \cite{bir,car} as well as opens up a
possibility to increase the efficiency of beam halo cleaning at
the stage of LHC luminosity upgrade \cite{bir}. Next, the
radiation of channeling particles, in that number, in crystal
undulators \cite{kap,bar,bel,kor}, allows to generate bright
gamma-beams. Both planar and axial channeling can be also used for
positron radiative cooling \cite{ugg}.

\begin{figure}[!ht]
\vspace{-7mm} \centering \psfull
    \epsfig{file=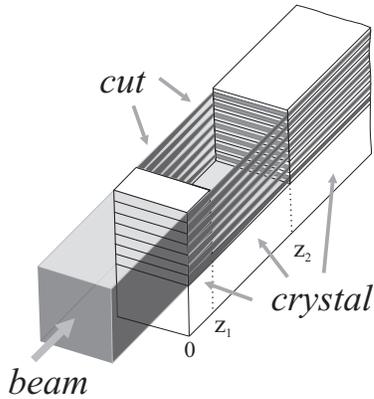,width=7cm}
\vspace{-38mm}
    \caption{A narrow cut between the planes $z=z_1$ and $z=z_2$,
which preserves a single crystal structure in front and behind
itself, allows to considerably reduce the proton transverse
energy.}
\end{figure}

The efficiency of all these applications is restricted by an
incomplete beam capture into the regime of stable channeling
motion, the probability of which hardly exceeds 85\% even for zero
beam divergence because of the fast dechanneling of, at least,
15\% of protons (positrons, ions which we will not mention further
for short) most intensively scattered by nuclei. To change the
situation and to reach a 99\% efficiency of proton capture into
the stable channeling motion we suggest in this letter to use a
narrow plane cut near the crystal entrance surface shown in Fig.
1. Similar cuts, widely used in channel-cut monochromators,
preserve a single crystal structure on the cut opposite sides. In
other words, atomic planes behind the cut are direct continuations
of that ones in front of it.

\vspace{5mm}
\begin{figure}[!ht]
\centering \psfull
    \epsfig{file=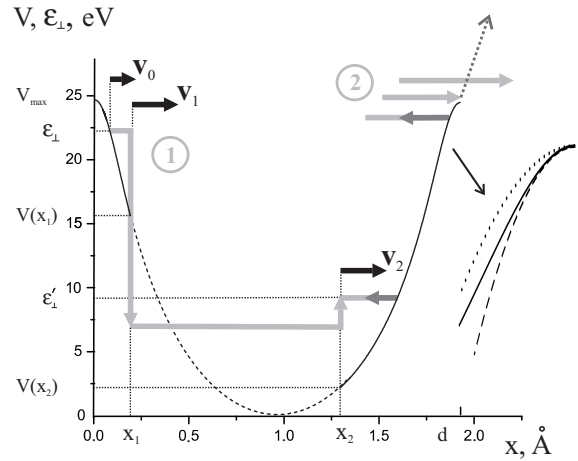,width=7.5cm}
    \caption{Dependencies of the averaged planar potential (thin black
curves) and transverse proton energy (thick gray lines) on the
transverse coordinate, 1 -- transverse energy reduction by the
crystal cut region (dashed curve), 2 -- dechanneling peak
formation by the protons with $\varepsilon_\perp \simeq V_{max}$.
Two extreme approximations to the potential near its maximum are
shown on the right.}
\end{figure}

The protons which start their transverse motion from the regions
of high nuclear number density, such as $|x| \lesssim 0.2 \AA$ and
$|x-d| \lesssim 0.2\AA$ in Figs. 2 and 3a, efficiently dechannel
at any, even zero beam divergence, since such protons recur to the
regions of high nuclear density -- see Fig. 3d. The cut prevents
the recurrence by reducing transverse energies of most of the
ready to dechannel protons. At first, passing through the crystal
plate $0 \leq z \leq z_1$ protons acquire transverse velocities
$v_1=v_x(z_1)$ directed to the channel center $x=d/2$, namely $v_1
> 0$ at $0 < x_0 < d/2$ and $v_1 < 0$ at $d/2 < x_0 <
d$, where $x_0=x(z=0)$, -- see Fig. 3b. Next, passing through the
cut, they drift towards the channel center (see Fig. 3c). As a
result, entering the crystal behind the cut the protons acquire
smaller potential energy $V(x_2)$ than the potential energy
$V(x_1)$ they lose at the cut entrance (here $x_{1,2} =
x(z_{1,2}))$. Moving further with the reduced transverse energy
most of the protons are now not able to reach the regions $|x|
\lesssim 0.2 \AA$ and $|x-d| \lesssim 0.2\AA$ of high nuclear
density (see Fig. 3e) from which they would dechannel in the cut
absence.

Of course, such a transverse energy reduction is possible if the
proton incidence angle $\vartheta_0 =v_0/ v_\parallel$ is
sufficiently small. Here $v_0=v_x(z=0)$ and $v_\parallel$ are,
respectively, the initial transverse and average longitudinal
velocity components. We will retain the later in the formulas to
make them valid both in relativistic and nonrelativistic cases. In
order to estimate the necessary $\vartheta_0$ value one can,
first, conjecture that to make a transverse kinetic energy
$\varepsilon v_1^2/2$ sufficiently small (say, less than a quarter
of the planar potential amplitude $V_{max}$, reserving another
quarter for the potential energy), the angle $\vartheta_1 =v_1/
v_\parallel$ should not exceed a half of the critical channeling
angle $\theta_{ch}=\sqrt{2 V_{max}/\varepsilon}$, where
$\varepsilon$ is the total proton energy. Second, in order not to
considerably change the angle $\vartheta_1$, at least, its sign,
the incidence angle $\vartheta_0$ should be considerably (say,
twice) less than $\vartheta_1$. Combining these two arguments one
comes to the condition
\begin {equation}
\vartheta_0 \lesssim \theta_{ch}/4
\end{equation}
of efficient dechanneling probability reduction by the cut.

To obtain Eq. (1) more rigorously let us develop a simple model of
the transverse energy reduction by the cut. Consider a proton
entering a crystal with a transverse velocity $v_0$ in a point
$x_0$ located in the high nuclear density region. Such a proton
possesses a transverse energy $\varepsilon_\bot=\varepsilon
v_0^2/2+V(x_0)$ close to $V_{max}$ -- see Fig. 2. After passing
through the crystal plate separated by the cut, the proton
acquires a coordinate $x_1$ and a velocity $v_1$. Entering then
the cut, it loses the potential energy $V(x_1)$, passes freely
through the cut with the constant velocity $v_1$, and acquires a
transverse coordinate
\begin {equation}
x_2 \equiv x(z_2)=x_1+v_1\cdot(t_2-t_1),
\end{equation}
where $t_1=z_1/v_\parallel$ and $t_2=z_2/v_\parallel$, and an
energy
\begin {equation}
\varepsilon_\bot^\prime \equiv \varepsilon_\bot(x_2)=\varepsilon
\frac{v_1^2}{2}+V(x_2)
\end{equation}
at the second entrance to the crystal at $z=z_2$.

\begin{figure}[!ht]
\centering \psfull \vspace{-25mm}
    \epsfig{file=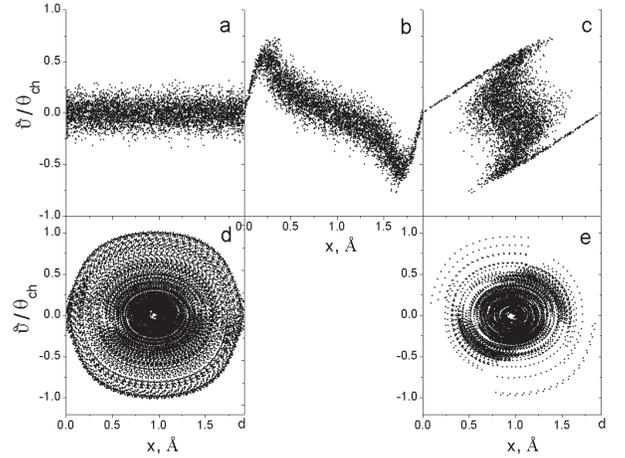,width=9.5cm}
\vspace{-35mm}
    \caption{Proton "phase space" a) at $z=0$, b) at $z=z_1$, c) at $z=z_2$ and e) at
$z=z_2+\pi v_\parallel /2 \omega$ in the cut presence and d) at
$z=z_2+\pi v_\parallel /\omega$ in its absence.}
\end{figure}

To evaluate the cut coordinates $z_{1,2}$ which minimize the
transverse energy $\varepsilon_\bot^\prime$ analytically let us
use the parabolic approximation
\begin {equation}
V(x)=\frac{k}{2}\left(x-\frac{d}{2}\right)^2
\end{equation}
correctly describing the motion of most part of protons in the
planar potential at small incidence angles. Substituting the
corresponding x-coordinate
\begin {equation}
x_1 = d/2 + a_0 \cos \omega t_1 + v_0 \sin \omega t_1
\end{equation}
and velocity x-component
\begin {equation}
v_1 = - a_0 \omega \sin \omega t_1 + v_0 \omega \cos \omega t_1
\end{equation}
one can represent $\varepsilon_\bot^\prime$ in the form of a
bilinear combination of $v_0$ and $a_0$ with some coefficients
depending on $z_1$ and $z_2$. Here $a_0=x_0-d/2$ and
$\omega=\sqrt{k/\varepsilon}$ are, respectively, the amplitude and
angular frequency of the channeling oscillations. The coefficient
by the product $a_0 v_0$ vanishes and Eq.(3) takes the form
\begin {equation}
\varepsilon_\bot^\prime = \varepsilon \frac{v_0^2}{2}\cot^2 \omega
t_1 + k \frac{a^2_0}{2}\tan^2 \omega t_1
\end{equation}
if the cut coordinates satisfy the relation
\begin {equation}
z_2=z_1+\frac{2 \cot (2\omega t_1) v_\parallel}{\omega}.
\end{equation}
We find this choice of the coordinate $z_2$ quite appropriate
since Eq. (7) is much simpler than the explicit form of Eq. (3),
it satisfies the symmetry conditions $\varepsilon_\bot^\prime
(a_0) = \varepsilon_\bot^\prime (-a_0)$ and
$\varepsilon_\bot^\prime (v_0) = \varepsilon_\bot^\prime (-v_0)$
and reaches a minimum both at $a_0=0$ and $v_0=0$. In addition,
Eq. (7) allows to find the coordinate
\begin {equation}
z_1=\frac{v_\parallel}{\omega}\arctan
\sqrt{\frac{\vartheta_0}{\theta (a_0)}},
\end{equation}
at which the minimal value
\begin {equation}
\varepsilon_\bot^\prime = 2 \frac{\vartheta_0}{\theta (a_0)}
V(a_0)
\end{equation}
of (7) is reached. Here $V(a_0)=k a_0^2/2$ is the initial
potential energy and $\theta (a_0) = (1/v_\parallel)\sqrt{2
V(a_0)/\varepsilon}$ is the proton deflection angle. Putting
$a_0=d/2$ one obtains $V(d/2) \simeq V_{max}$ and $\theta(d/2)
\simeq \theta_{ch}$. Assuming that protons quickly dechannel at
$\varepsilon_\bot^\prime
> \varepsilon_{\bot dech}$, where $\varepsilon_{\bot dech}$
is an effective dechanneling energy, one indeed comes to the
condition (1) of efficient cut use at quite natural value of
$\varepsilon_{\bot dech} =V_{max}/2$.

From Eqs. (8) and (9) one easily obtains that
\begin {equation}
\frac{z_2}{z_1}= 1+\frac{\sqrt{\vartheta_0/
\theta_{ch}}}{\arctan\sqrt{\vartheta_0/
\theta_{ch}}}\left(\frac{\theta_{ch}}{\vartheta_0} - 1\right)
\end{equation}
at $\theta(d/2) \simeq \theta_{ch}$ and $z_2/z_1 \simeq 4.235$ at
$\vartheta_0=0.25 \theta_{ch}$. Putting $k=8V_{max}/d^2 \simeq
53.1 eV/\AA^2$ one can easily estimate that $z_1$ value increases
from $100 \AA$ at $\varepsilon = 1 MeV$, to $0.2 \mu m$ at
$\varepsilon = 1 GeV$ and to $17 \mu m$ at the LHC energy of
$\varepsilon = 7 TeV$. Expressing the equation of transverse
motion in the form $\varepsilon v_\parallel^2 d^2x/d^2z =
-V(x)^\prime$ and observing that $z_{1,2} \propto v_\parallel
\sqrt{\varepsilon}$, one can easily see that the
$\varepsilon_\bot$ evolution from $z=0$ to $z = z_2$ remains the
same for any energy.
\begin{figure}[!ht]
\centering \psfull
    \epsfig{file=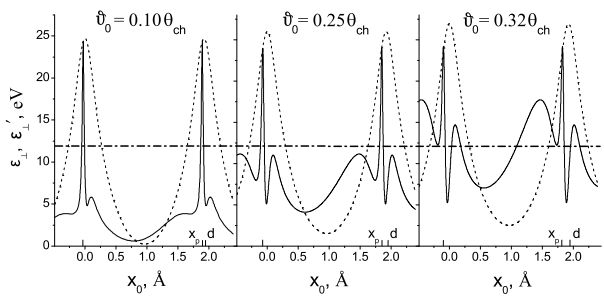,width=9cm}
\vspace{-6mm}
    \caption{Dependencies of the proton transverse energy at $z=0$
(dashed) and $z=z_2$ (solid curves) on the initial proton
coordinate at different angles of proton incidence with respect to
the (110) $Si$ plane.}
\end{figure}

The results of such an evolution simulated for three incidence
angles $\vartheta_0$ using the Moliere approximation for the
averaged planar potential are illustrated by Fig. 4. A comparison
of the solid and dashed curves demonstrates an efficient
$\varepsilon_\bot^\prime$ reduction by the cut in the regions of
high nuclear density. At $\vartheta_0 < \theta_{ch}/4$ this
reduction allows to reach $\varepsilon_\bot^\prime < V_{max}/2$
and, thus, to avoid fast dechanneling by 98-99\% of the protons
while the rest 1-2\% of them form sharp peaks at $x_0 = x_p$ and
quickly dechannel. Though the origin of these "dechanneling peaks"
can not be explained using the approximation (4), Eqs. (7) and
(10) correctly describe the $\vartheta_0$ dependence of the
maximum $\varepsilon_\bot^\prime$ value reached outside the
peaks. In particular, one can see that $\varepsilon_\bot^\prime$
indeed reaches $V_{max}/2 \simeq 12.2 eV$ at $\vartheta_0$
slightly exceeding $\theta_{ch}/4$.

The dechanneling peaks originate from the protons with $x_0 \simeq
x_p$ and $\varepsilon_\bot \simeq \varepsilon v_0^2 /2+V(x_p)
\simeq V_{max}$ which slow down their transverse motion near the
potential maxima acquiring very low velocities $v_1$ and
coordinates $x_1 \simeq nd, n=0, \pm1, \pm2,..$, at the cur
entrance. Their further motion with very low transverse
velocities inside the cut preserves the proton transverse
positions $x_2 \simeq x_1 = nd$ near the potential maxima. As a
result, entering the crystal behind the cut such protons acquire
transverse energies $\varepsilon_\bot^\prime \simeq V(x_2) \simeq
V_{max}$ corresponding to the dechanneling peak height. A few
outermost trajectories in Fig. 3e belong to such protons.

A small width of the dechanneling peaks takes its origin from the
high potential curvature $k^\prime$ near its maxima due to which
even a slight $x_0$ deviation from the peak positions $x_p$ leads
either to the fast proton reflection from the potential wall at
$\varepsilon_\bot < V_{max}$ or to the passage above the potential
maximum at $\varepsilon_\bot > V_{max}$ -- see Fig. 2. In both
cases the protons quickly joint the process of transverse energy
reduction illustrated by Figs. 3 b, c and e.

In order to estimate the fraction of protons contributing to the
dechanneling peaks we consider an analytical model of proton
motion in the potential $V(x) \simeq V(0)-k^\prime x^2/2$
approximating the realistic one near its maximum. Substituting
corresponding x-coordinate
\begin {equation}
x_1 = x_0 \coth \lambda t_1 + v_0 \sinh \lambda t_1,
\end{equation}
where $\lambda=\sqrt{k^\prime /\varepsilon}$, and velocity
$v_1=dx(t_1)/dt$ into Eqs. (2) and (3) one obtains
\begin {equation}
\varepsilon_\bot^\prime (x_0) \simeq V_{max} + \frac{1}{2}
\frac{\partial^2 \varepsilon_\bot^\prime (x_0)}{\partial^2
x_0}\Biggr|_{x_0=x_p} (x_0-x_p)^2,
\end{equation}
where $x_p=x_p(v_0)$ is the dechanneling peak coordinate fixed by
the equation $\partial \varepsilon_\bot^\prime (x_0)/\partial x_0
= 0$ and
\begin {eqnarray}{l}
\nonumber \frac{\partial^2 \varepsilon_\bot^\prime
(x_0)}{\partial^2 x_0}\Biggr|_{x_0=x_p} =-1-\lambda(t_2-t_1)
\\ \times [\lambda (t_2-t_1) \sinh^2 \lambda t_1 + 2 \coth \lambda t_1 \sinh
\lambda t_1]
\end{eqnarray}
is the second $\varepsilon_\bot^\prime$ derivative which directly
determines the dechanneling peak width $\triangle x_p$ at the
level of a dechanneling energy $\varepsilon_{\bot dech}$. This
width allows to evaluate the relative number $\triangle x_p/d$ of
protons with $\varepsilon_\bot
> \varepsilon_{\bot dech}$ which we will use as an estimate
\begin {equation}
P_{dech} = 2 \sqrt{\frac{2(V_{max}- \varepsilon_{\bot dech})}
{\partial^2 \varepsilon_\bot^\prime (x_0)/\partial^2 x_0
\Bigr|_{x_0=x_p} }}
\end{equation}
of the dechanneling probability. To evaluate the later one has to
substitute the potential curvature $k^\prime$ the choice of which
is somewhat ambiguous, as an insert on the right in Fig. 2
demonstrates. Its lower estimate $k^\prime_{min} =E_{max}/x_{in}
\simeq 475 eV/\AA^2$ follows from the values of the maximum
electric field strength $E_{max} \simeq 6.17 GeV/cm$ and
inflection point x-coordinate $x_{in} \simeq 0.13\AA$, while its
maximum value $k^\prime_{max} =d^2V(x)/d^2 x|_{x=0} \simeq 846
eV/\AA^2$ is reached in the point $x=0$ of potential maximum. At
$\varepsilon_{\bot dech} =V_{max}/2$ the probability (15) is equal
to 3.24\% at $k^\prime = k^\prime_{min}$ and to 1.24\% at
$k^\prime = k^\prime_{max}$. The averaged value 2.24\% of these
estimates is in a good agreement with the values 2.13\% and 2.59\%
obtained by simulations for the cases of, respectively,
$\vartheta_0 =0$ and $\vartheta_0= 0.5 \mu rad$. The probability
(15) will be even lower if the protons effectively dechannel in a
relatively short crystal at higher $\varepsilon_{\bot dech}$
values, for instance, at $\varepsilon_{\bot dech} =3V_{max}/4$.
Indeed, Eq. (15) predicts in this case that $P_{dech} = 2.29\%$
and 0.91\%, respectively, at $k^\prime = k^\prime_{min}$ and
$k^\prime = k^\prime_{max}$ while the simulations give 1.13\% at
$\vartheta_0 =0$ and 1.34\% at $\vartheta_0 = 0.5 \mu rad$.

Thus, the cut reduces the transverse energy of 98-99\% of protons
of a beam with divergence (1) down to the level securing their
stable channeling motion. The rest 1-2\% of the protons avoid
efficient energy reduction and dechannel fast. This picture
manifests itself independently of the proton energy provided the
cut coordinates which are, naturally, energy dependent, are chosen
according to Eqs. (8) and (9).  Thus, to illustrate the general
features of the cut functioning by Monte Carlo simulations any
proton energy may be chosen. Let us therefore consider the LHC
proton deflection by bent crystals as the most challenging
channeling application.

\begin{figure}[!ht]
\centering \psfull \vspace{-10mm}
    \epsfig{file=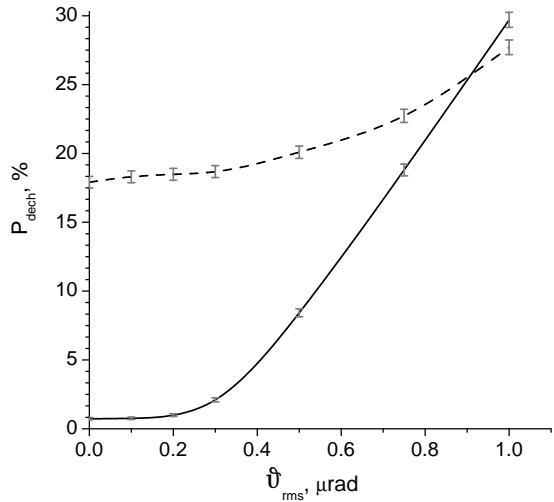,width=9cm}
\vspace{-50mm}
    \caption{Dependence of the 7 TeV proton dechanneling probability
in a 1 cm bent $Si$ crystal with (solid) and without (dashed
curve) the cut on the r.m.s. incidence angle.}
\end{figure}

In previous deflection experiments the low capture probability was
predetermined simultaneously  by the high incident beam
divergence, relatively thick amorphous near-surface crystal layer
and fast dechaneling of protons moving in the regions of high
nuclear density. At low capture probability a high deflection
efficiency could be realized only in the multi-pass mode
\cite{bir,car}. However the divergence of the beam fraction
encountering a crystal placed in the beam halo of a high energy
proton accelerator can, in fact, be made very low \cite{her,kle}.
The amorphous layer also can be made as thin as $0.01 \mu m$
\cite{bir}. In the circumstances the fast dechaneling becomes the
only obstacle impeding an efficient single pass proton deflection
and since the cut drastically reduces the number of quickly
dechaneling protons, its use has to make an efficient single pass
deflection possible.

To demonstrate this we simulated the $7 TeV$ LHC proton deflection
by a 1 cm $Si$ crystal with 100m bending radius and cut
coordinates $z_1=17\mu m$ and $z_2=71\mu m$ evaluated using Eqs.
(8) and (9). To make the simulation process more adequate and
efficient we treated the large-angle single Coulomb scattering
separately from the small-angle multiple one. The angular
distribution of protons in the incident beam was assumed to be
both Gaussian and cylindrically symmetric. The simulated
dependence of the dechanneling fraction on the root mean square
(r.m.s.) divergence angle $\vartheta_{rms}$ is given in Fig. 5. On
the average the cut begins to suppress the dechanneling process
starting from $\vartheta_{rms} \simeq 0.9 \mu rad \simeq 0.34
\theta_{ch}$ while the highest efficiency of its use is reached at
$\vartheta_{rms} \leq 0.3 \mu rad \simeq 0.1 \theta_{ch}$ when the
dechanneling probability decreases down to 1\% as predicted by the
analytical model.

\begin{figure}[!ht]
\centering \psfull \vspace{-3mm}
    \epsfig{file=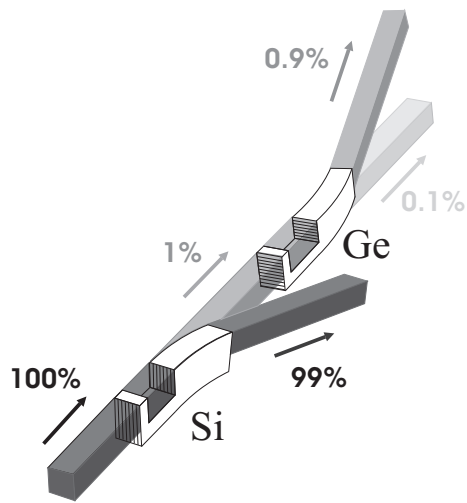,width=7cm}
\vspace{-25mm}
    \caption{A pair
of 1 cm $Si$ and $Ge$ crystals with cuts bent in orthogonal planes
with 100m bending radii allows to reach a 99.9\% single-pass
proton deflection efficiency at the LHC.}
\end{figure}

The simulations also show that the protons dechanneled in the
crystal which is bent in horizontal xz plane acquires a r.m.s.
divergence angle of about $0.7 \mu rad$ in the vertical yz plane.
Such a divergence allows to use a second crystal bent in the yz
plane to deflect about 90\% of these protons, as is shown in Fig.
6. Though the high beam divergence impedes the dechanneling
suppression by the cut, a combined single-pass deflection
efficiency by a pair of $Si$ and $Ge$ crystals with cuts
approaches 99.9\%. Both the single passage though the crystal and
the proton moving off the high nuclear density regions will
drastically diminish the rate of nuclear reactions accompanying
the proton deflection by crystals with cuts.

Thus, to reach a really high single-pass deflection efficiency the
beam divergence angle has not to exceed much $0.1 \mu rad$.  The
angle which the proton beam axis forms with the crystal planes
must be of the same order of value. Since the best goniometers
allow to position a crystal only with a microradian accuracy, the
necessary beam alignment along the crystal planes can most easily
be reached by a proton deflection by a magnetic field with the
typical integral value of only $0.01 T \cdot m$ which should be
adjusted to maximize the deflection probability.

The symmetry of the cut region from Fig. 1 prevents the
dusturbance of  $\varepsilon_\bot$ reduction process by a
distortion accompanying the cut formation. However the crystal
bending in the horizontal plane violates the symmetry and can
distort the cut region upsetting the $\varepsilon_\bot$ reduction
process. The simplest possibility to avoid such a distortion is to
leave the cut region straight, starting the crystal bending
at $z > z_2$. To prevent the distortion of the cut region if it is
nevertheless bent we suggest to tightly fix a crystal plane or set
of planes with random crystal structure orientation and elastic
properties identical to that of the removed crystal plane inside
the cut. The method of high-resolution x-ray scattering
\cite{piet} can be used to check both the crystal plane matching
and alignment on the opposite sides of the cut.

Note briefly that the reduction of both the dechanneling
probability and averaged transverse energy by the cut fabricated
in the crystals used to generate either usual channeling radiation
or channelled positron radiation from crystal undulators
\cite{kap,bar,bel,kor} will diminish the intensity of radiation
from non channeled positrons, make the radiation spectrum more
narrow, increase the effectively used crystal length and reduce
the rate of positron interaction with nuclei.

The positron transverse energy reduction by the cut also allows to
increase the number of positrons involved in the process of
radiative cooling being effective only in the regions of
negligible nuclear density \cite{ugg}. Note that the cut can be
used to improve the efficiency of radiative cooling both in the
planar and axial cases. At that the two-dimensional distribution
of both the nuclear density and averaged potential allows to reach
even higher reduction of the quickly dechaneling proton or
positron fraction in the axial case than in the planar one.

Note also that the cut application to the case of negatively
charged particles will, on the opposite, several times and
more increase the dechanneling and nuclear reaction probabilities
both in planar and axial cases.

In principle, dechanneling probability reduction by the cut can
both increase the depth and modify the profile of non relativistic
ion implantation. However the small lengths $z_{1,2} \sim 10-100
nm$ will seriously complicate the cut production. A crystalline
surface layer with low average atomic number can be used instead
of such narrow cuts. The crystal planes of this layer also have to
be the direct continuations of that in the crystal. A vast
experience in heterostructure growth \cite{alf} allows one to
argue that a high quality $GaAl$ layer can be produced on a $Ge$
crystal most readily.

To summarize, the fabrication of a narrow plane cut near the
crystal surface allows to ten times and more decrease both the
dechanneling probability and nuclear reaction rate of positively
charged particles in crystals. This possibility can many times
increase the efficiency of both the beam deflection and halo
cleaning at proton and ion colliders, of the gamma-quantum
emission by positrons in crystal undulators and of the positron
beam cooling.

The author gratefully acknowledges useful discussions with V.G.
Baryshevsky, X. Artru, I.D. Feranchuk, V.A. Maisheev, N.A.
Poklonsky,  A.P. Ul'yanenko and I.A. Yazynin. This work was partly
supported by the \# 03-52-6155 INTAS Project.

\end{document}